\documentclass[prd,aps,a4paper,twocolumn,eqsecnum,nofootinbib,floatfix]{revtex4}  %

\newif\ifusesec
\usesectrue  
   
\usepackage{graphicx} 
\usepackage{amsmath,amsfonts,amssymb}
\usepackage{appendix}
\usepackage{mathtools}
\usepackage[caption=false,justification=justified]{subfig}
\usepackage{xcolor}

\newcommand{\beq}{\begin{equation}}
\newcommand{\eeq}{\end{equation}}
\newcommand{\bea}{\begin{eqnarray}}
\newcommand{\eea}{\end{eqnarray}}

\begin{document}

\title{Gravitational waveform from radial infall at the third-and-half Post-Newtonian order}

\author{Giorgio Di Russo$^{1}$, Donato Bini$^{1}$}
  \affiliation{
$^1$School of Fundamental Physics and Mathematical Sciences, Hangzhou Institute for Advanced Study, UCAS, Hangzhou 310024, China
$^2$Istituto per le Applicazioni del Calcolo ``M. Picone,'' CNR, I-00185 Rome, Italy\\
}

\date{\today}

\begin{abstract}
We compute the gravitational waveform associated with a radially infalling particle in a Schwarzschild black hole working in the center-of-mass system and in a post-Newtonian (PN) approximation. Our results reach the highest accuracy level fully displayed in the literature, namely  the 3.5PN order. The latter accuracy includes both conservative and  radiation-reaction contributions (at 2.5PN and 3.5PN) in the two-body dynamics, and corresponding effects in the waveform too. The apparent simplicity of the radial fall (namely, the 1-dimensional motion) contrasts with the peculiarity of the process which will end necessarily with the capture of the particle by the black hole, featuring strong field effects. In other words, our analysis being limited to the region of validity of the PN approximation, cannot capture (by definition of PN approximation) the final phase of the fall, but offers significant insights anyway. 
\end{abstract}

\maketitle

\section{Introduction}
\label{Intro}

The gravitational two-body problem (with all the associated aspects: conservative, radiation-reacted, local,  nonlocal) is currently being studied by using various
theoretical approaches and approximation schemes: post-Newtonian
(PN) \cite{Blanchet:2013haa},  post-Minkowskian (PM) \cite{Porto:2016pyg}, Multipolar Post-Minkowskian (MPM), PN-matched MPM, self-force (SF) \cite{Barack:2018yvs}, Effective-Field-Theory (EFT) \cite{Travaglini:2022uwo,Buonanno:2022pgc,Brunello:2025eso}, Effective One-Body (EOB) \cite{Buonanno:1998gg,Buonanno:2000ef}, Tutti Frutti (TF) \cite{Bini:2020nsb} (quoting mainly reviews for convenience).

Such approaches are all of perturbative type, and the inclusion in our computations of high-perturbative orders is required to obtain a more and more accurate description of the two-body dynamics.  In turn, this is necessary  to build accurate enough templates to be used in the analysis of data coming from gravitational wave Earth-based detectors. However, high-perturbative orders are then associated with additional (often non trivial) technical difficulties to be handled with. This is the case, for example, of considering radiation-reaction effects which carry most of the present bottlenecks.

The current state-of-the-art in the  (PN-matched) MPM  formalism  is the 4.5PN accuracy, but only in the phasing of quasi-circular inspiralling binaries
\cite{Blanchet:2022vsm,Blanchet:2023bwj,Blanchet:2023sbv}. Out of quasi-circular motion the known accuracy is 3.5PN, i.e. the level considered in the present paper. Genuine PM formalism, instead, is currently completing results at the $O(G^5)$ order.  

Here, within the above mentioned PN-matched MPM formalism, we will  
compute the waveform $ W =\lim_{r\to \infty}\frac{r}{4G} \bar m^{i} \bar m^{j }  h_{i j}$ emitted by radially infalling orbits (for which pioneering studies date back to the 70s \cite{Zerilli:1970wzz,Davis:1971gg,Davis:1972ud,Ruffini:1973ky}) at the 3.5PN, and display the explicit values of both the time-domain and the frequency-domain waveform. 

We use the mostly positive metric convention ($-+++$), and denote the PN expansion parameter as $\eta=\frac{1}{c}$.
As it is customary in the MPM formalism,  we will express all source and gauge multipole moments (see below) in terms of the dynamical variables of the binary system in  \lq\lq modified  harmonic coordinates," eee  Ref. \cite{Blanchet:2013haa}.

\section{The complete waveform within the MPM formalism and at the 3.5PN accuracy level}

The  transverse-traceless (TT) asymptotic waveform  $\lim_{r\to \infty}( r \, h_{ij}^{\rm TT})$ is conveniently
summarized in the complex (metric) quantity
\bea
h_c(t_r,\theta,\phi) &=&\lim_{r\to \infty}(r( h_+ -  i h_\times))\nonumber\\
&=&\lim_{r\to \infty} \bar m^{\mu } \bar m^{\nu }r\, h_{\mu \nu}\,,
\eea
where 
\beq \label{tret}
t_r = t-\frac{r}{c}-\frac{2 G{\cal M} }{c^3}\ln \left( \frac{r}{b_0}\right),
\eeq
is a Bondi-type  retarded time variable (depending on an
arbitrary length scale, $b_0$);  $\bar m^{\mu }$ is a null polarization vector, chosen so 
to have only center-of-mass (cm) spatial components $\bar m^j$, see Refs. \cite{Bini:2023fiz,Bini:2024rsy,Bini:2024ijq}, orthogonal to the observation direction denoted by $n^i$.
The cm four velocity $U$ is aligned with the time direction of a Lorentz frame, $e_0=U$.
In the cm local-rest-space  we define either a spatial (with respect to $U$) triad $(n,e_{\hat\theta},e_{\hat \phi})$
\bea
\label{spat_triad}
n^{i}&=& (s_\theta c_\phi, s_\theta s_\phi, c_\theta)\,,\nonumber\\
e_{\hat \theta}^i&=&\frac{\partial n^{i}}{\partial \theta}\,,\nonumber\\
e_{\hat \phi}^i&=&\frac{1}{s_\theta}\frac{\partial n^{i}}{\partial \phi}\,,
\eea
or a (complex, orthogonal to $U$) null triad $n, m, \bar m$, with
\bea
m^i&=&\frac{1}{\sqrt{2}}(e_{\hat \theta}^i+i e_{\hat \phi}^i)\,,\nonumber\\
\bar{m}^i&=&\frac{1}{\sqrt{2}}(e_{\hat \theta}^i-i e_{\hat \phi}^i)\,,
\eea
where the normalization factor $1/\sqrt{2}$ is arbitrary and chosen so that $m \cdot \bar m =+1$.
We used the compact notation $[\cos \alpha, \sin\alpha]=[c_\alpha,s_\alpha]$, and we assumed a system of Cartesian-like coordinates with origin in the cm of the system. 
Explicitly, the components of $\bar{m}^i$ read
\beq
\bar{m}^i {=} \frac{1}{\sqrt{2}}\left((c_\theta c_\phi{+}i s_\phi),(c_\theta s_\phi{-}i c_\phi),{-}s_\theta\right)\,.
\eeq

Note that as soon as the two bodies start interacting and undergo radiation-reaction effects the cm recoils, (i.e., it accelerates and moves because of the consequent loss of linear momentum starting from the 3.5PN accuracy). Therefore, when studying cm motion effects, one usually refer quantities to an \lq\lq incoming cm" in the far past.

The PN-matched MPM formalism computes  the  multipolar decomposition  of $h_c(t_r,\theta,\phi)$ in terms of the (complex, time-domain) waveform $W$, namely  
 \beq
 h_c(t_r,\theta,\phi) \equiv 4 G \eta^4  W(t_r,\theta,\phi)\,,
 \eeq
decomposed in electric-type ($U_L$) or magnetic type ($V_L$) multipoles as follows
\bea
\label{W_deco}
W(t_r,\theta,\phi)&=& U_2+ \eta (V_2 +U_3) + \eta^2 (V_3+U_4)\nonumber\\ 
&+& \eta^3 (V_4+U_5)+ \cdots \,.
\eea
Here, each $U_\ell$ denotes an even-parity $2^\ell$ radiative multipole contribution and similarly $V_\ell$ denotes an odd-parity $2^\ell$ radiative multipole.
In the MPM formalism $U_\ell$ and $V_\ell$ are expressed in terms 
of symmetric-trace-free (STF) Cartesian-type tensors of order $\ell$ (the radiative multipole moments  $U_{ i_1 i_2 \cdots i_{\ell}}(t_r)$
and  $V_{ i_1 i_2 \cdots i_{\ell}}(t_r)$) according to the following definitions
\bea
\label{UVdefs}
U_\ell(t_r,\theta,\phi) &=& \frac{1}{\ell!} \bar m^{i} \bar m^{j } n^{i_1} n^{i_2} \cdots n^{i_{\ell-2}} U_{i j i_1 i_2 \cdots i_{\ell-2}}(t_r)\,, \nonumber\\
V_\ell(t_r,\theta,\phi) &=& - \frac{1}{\ell!}\frac{2\ell}{\ell+1} \bar m^{i} \bar m^{j } n^c  n^{i_1} n^{i_2} \cdots n^{i_{\ell-2}}\times \nonumber\\
&&  \epsilon_{cd i}V_{j d i_1 i_2 \cdots i_{\ell-2}}(t_r)\,.
\eea
The PN-matched MPM formalism computes the  radiative moments by iteratively solving  Einstein's vacuum equations in the exterior zone. In the end, the radiative moments are expressed  as nonlinear retarded functionals of two other types of multipolar moments: (i) the \lq\lq source moments," 
\beq
 I_{L}(t_r)\,,\qquad J_{L}(t_r)\,,
\eeq 
with $L=i_1 i_2 \cdots i_{\ell}$ a multi-index notation,
together with various \lq\lq gauge moments":  
\bea
&& W_{L}(t_r)\,,\quad X_{L}(t_r)\,,\quad Y_{L}(t_r)\,,\quad  Z_{L}(t_r)\,.
\eea
The source and gauge moments are then computed  as explicit expressions of the positions and velocities of the (cm-relative) motion. 

In this work, we shall compute all the needed multipole moments to reach the 3.5PN accuracy level in the incoming (i.e., at the beginning of the radial fall) cm frame of the system, by using  modified harmonic coordinates \cite{Blanchet:2013haa}.

The frequency-domain waveform $W(\omega,  \theta,\phi) $ is then obtained by Fourier-transforming (over the retarded time variable $t_r$) 
the radiative moments,
\bea
U_{L}(\omega)= \int_{- \infty}^{+ \infty} dt_r e^{i \omega t_r} U_{L}(t_r)\,, \nonumber \\
V_{L}(\omega)= \int_{- \infty}^{+ \infty} dt_r e^{i \omega t_r} V_{L}(t_r)\,.
\eea
This  leads to the following, explicit multipolar expansion of the frequency-domain complex waveform 
$h_c(\omega,\theta,\phi) \equiv 4 G \eta^4 W(\omega,  \theta,\phi)$,
\bea
\label{Wom_th_phi}
W(\omega,  \theta,\phi)  &\equiv& U_2+\eta (V_2  +U_3 )\nonumber\\ 
&+& \eta^2 (V_3 +U_4 )+\eta^3 (V_4 +U_5 )\nonumber\\
&+& \eta^4 (V_5 +U_6 )+\eta^5 (V_6 +U_7 )\nonumber\\
&+& \eta^6 (V_7 +U_8 )+ \eta^7 (V_8 +U_9 )\nonumber\\
&+& O(\eta^8)\,,
\eea
where we have omitted the variable dependence of all quantities at the right-hand-side (i.e.,  $U_2=U_2(\omega,\theta,\phi)$, etc.) for simplicity. Electric-type (up to $l=9$)  and magnetic-type (up to $l=8$) multipoles are included in $W$ in order to reach a total 3.5PN accuracy for the waveform itself, as already stated. 

For example, the quadrupole contribution reads
\beq
U_2(\omega,\theta,\phi)=  \frac{1}{2!} \bar m^{i} \bar m^{j } U_{i j}(\omega)\,.
\eeq
To compute $U_2$ (and then $W$)  at the 3.5PN accuracy level, we recall that the radiative multipoles $U_L$ and $V_L$ should be expressed in terms of the so-called canonical multipoles $M_L$ and $S_L$. For example,
\bea
\label{U_def}
U_{ij} &=& M_{ij}^{(2)} + \eta^3 U_{ij}^{\rm 1.5PN}  +\eta^5 U_{ij}^{\rm 2.5PN}\nonumber\\
&+&\eta^6 U_{ij}^{\rm 3PN}+\eta^7 U_{ij}^{\rm 3.5 PN}+O\left(\eta^8\right)\,,\nonumber\\
U_{ijk}&=&M_{ijk}^{(3)}+\eta^3U_{ijk}^{\rm 1.5 PN}+\eta^5U_{ijk}^{\rm 2.5PN}\nonumber\\
&+&\eta^6U_{ijk}^{\rm 3PN}+O\left(\eta^7\right)\,,\nonumber\\
U_{ijkl}&=&M_{ijkl}^{(4)}+\eta^3 U_{ijkl}^{\rm 1.5 PN}+\eta^5 U_{ijkl}^{\rm 2.5PN}\nonumber\\
&+&O\left(\eta^6\right)\,,\nonumber\\
U_{ijklm}&=&M_{ijklm}^{(5)}+\eta^3 U_{ijklm}^{\rm 1.5PN}+O\left(\eta^5\right)\,,\nonumber\\
U_{ijklmn}&=&M_{ijklmn}^{(6)}+O\left(\eta^4\right)\,, \nonumber\\
U_{ijklmnp}&=&M_{ijklmnp}^{(7)}+O\left(\eta^3\right)\,, \nonumber\\
U_{ijklmnpq}&=&M_{ijklmnpq}^{(8)}+O\left(\eta^2\right)\,,\nonumber\\
U_{ijklmnpqr}&=&M_{ijklmnpqr}^{(9)}+O\left(\eta\right)\,.
\eea
Successively, the canonical multipoles $M_L$ and $S_L$ are expressed in terms of the source multipole moments $I_L$ and $J_L$ and the gauge moments $W_L$, $X_L$, $Y_L$, $Z_L$. For example, following the notation introduced in Ref. \cite{Bini:2026dvn}
\bea
\label{blocks_0}
M_{ij} &=& 
I_{ij}+ 4G \eta^5M_{ij}^{W_0\,I_2}\nonumber\\
&+&4G\eta^7\Big(M_{ij}^{W_2\,I_2}+M_{ij}^{Y_2\, I_2}+M_{ij}^{X_0\,I_2}+M_{ij}^{W_1\,I_3}\nonumber\\
&+&M_{ij}^{Y_1\, I_3}+M_{ij}^{W_0\,W_2}+M_{ij}^{W_1\,W_1}+M_{ij}^{W_0\, Y_2}+M_{ij}^{W_1\, Y_1}\nonumber\\
&+&M_{ij}^{Z_1\,I_2}+M_{ij}^{W_1\, J_2}+M_{ij}^{Y_1\, J_2}\Big) \,,
\eea
where
\bea
M_{ij}^{W_0\,I_2}&=&W^{(2)}I_{ij}-W^{(1)}I_{ij}^{(1)}\,,\nonumber\\
M_{ij}^{W_2\, I_2}&=&\frac{4}{7}W_{a\langle i}^{(1)}I_{j\rangle a}^{(3)}+\frac{6}{7}W_{a\langle i}I_{j\rangle a}^{(4)}\,,\nonumber\\
M_{ij}^{Y_2\, I_2}&=&-\frac{1}{7}Y_{a\langle i}^{(3)}I_{j\rangle a}-Y_{a\langle i}I_{j\rangle a}^{(3)}\,,\nonumber\\
M_{ij}^{X_0\, I_2}&=&-2XI_{ij}^{(3)}\,,\nonumber\\
M_{ij}^{W_1\, I_3}&=&-\frac{5}{21}W_a^{(4)}I_{ija}+\frac{1}{63}W_a^{(3)}I_{ija}^{(1)}\,,\nonumber\\
M_{ij}^{Y_1\,I_3}&=&-\frac{25}{21}Y_a^{(3)}I_{ija}-\frac{22}{63}Y_a^{(2)}I_{ija}^{(1)}+\frac{5}{63}Y_a^{(1)}I_{ija}^{(2)}\,,\nonumber\\
M_{ij}^{W_0\,W_2}&=&2W^{(3)}W_{ij}+2W^{(2)}W_{ij}^{(1)}\,,\nonumber\\
M_{ij}^{W_1\, W_1}&=&-\frac{4}{3}W_{\langle i}W_{j\rangle}^{(3)}\,,\nonumber\\
M_{ij}^{W_0\, Y_2}&=&2W^{(2)}Y_{ij}\,,\nonumber\\
M_{ij}^{W_1\,Y_1}&=&-4W_{\langle i}Y_{j\rangle}^{(2)}\,,\nonumber\\
M_{ij}^{Z_1\, I_2}&=&\epsilon_{ab\langle i}\Big[\frac{1}{3}I_{j\rangle a}Z_b^{(3)}-I_{j\rangle a}^{(3)}Z_b\Big]\,,\nonumber\\
M_{ij}^{Y_1\, J_2}&=&\epsilon_{ab\langle i}\Big[-\frac{4}{9}J_{j\rangle a}Y_b^{(2)}+\frac{8}{9}J_{j\rangle a}^{(1)}Y_b^{(1)}\Big]\,.
\eea
The other terms in Eq. \eqref{U_def} at our accuracy  read 
\bea
\label{lista_up_2.5pn}
U_{ij}^{\rm 1.5PN}(t_r) &=&  2 G {\cal M}  \int_0^{+\infty} d\tau\,  M_{ij}^{(4)}(t_r-\tau)\ln \left(\frac{\tau}{C_{I_2}}\right)\,,\nonumber\\
\eea
with  ${\cal M} \equiv \frac{E}{c^2}$ denoting the total ADM mass in the CM system, and
\beq
C_{I_2}=\frac{2 b_0 e^{-11/12}}{c}\,.
\eeq
Here $U_{ij}^{\rm 1.5PN}(t_r)$ needs to be computed at 1PN fractional accuracy, while the
additional, genuine 2.5PN contributions, see Eq. \eqref{U_def}, need to be computed at the Newtonian accuracy.  They are given by
\bea
U_{ij}^{{\rm 2.5PN}}(t_r) &=&  G  \left( 
U_{ij}^{{\rm 2.5PN}\rm (mem)}
+U_{ij}^{{\rm 2.5PN} I_2I_2}\right.\nonumber\\
&+&\left. U_{ij}^{ {\rm 2.5PN} I_2J_1}\right)\,,
\eea
where  
\bea
U_{ij}^{{\rm 2.5PN} \rm  (mem)}(t_r)&=&- \frac{2}{7} \int_0^{+\infty} \! d\tau\!\left[ M^{(3)}_{a\langle i} M^{(3)}_{j\rangle a}\right]\!(t_r-\tau)\,,\nonumber\\
U_{ij}^{{\rm 2.5PN}\, I_2I_2}(t_r)&=&  \frac{1}{7}\, M^{(5)}_{a\langle i} M^{}_{j\rangle a} - \frac{5}{7} \, M^{(4)}_{a\langle i} M^{(1)}_{j\rangle a}\nonumber\\
&-& \frac{2}{7}\, M^{(3)}_{a\langle i} M^{(2)}_{j\rangle a}  
\,,\nonumber\\
U_{ij}^{{\rm 2.5PN}I_2J_1}(t_r)&=& 0 \,.\nonumber\\
U_{ij}^{\rm 3 PN}(t_r)&=&2G^2{\mathcal M}^2\int_0^{\infty}d\tau M_{ij}^{(5)}(t_r-\tau)\nonumber\\
&\times&\Big[\log^2\Big(\frac{\tau}{2b_0}\Big)+\frac{11}{6}\log\Big(\frac{\tau}{2b_0}\Big)\nonumber\\
&-&\frac{107}{105}\log\Big(\frac{\tau}{2r_0}\Big)+\frac{124627}{44100}\Big]\,,\nonumber\\
U_{ij}^{\rm 3.5PN}(t_r)&=&G\Big[U_{ij}^{\rm 3.5PN(mem)}(t_r)+U_{ij}^{\rm 3.5PN\,I_2I_4}(t_r)\nonumber\\
&+&U_{ij}^{\rm 3.5 PN\, I_3I_3}(t_r)+U_{ij}^{\rm 3.5 PN\, J_1J_3}(t_r)\nonumber\\
&+&U_{ij}^{\rm 3.5PN\,J_2J_2}(t_r)+U_{ij}^{\rm 3.5PN\, J_3I_2}(t_r)\nonumber\\
&+&U_{ij}^{\rm 3.5PN\,I_3J_2}(t_r)\Big]\,,
\eea
where
\begin{widetext}
\bea
U_{ij}^{\rm 3.5PN(mem)}(t_r)&=&\int_0^\infty d\tau\Big[-\frac{5}{756}M_{ab}^{(4)}M_{ijab}^{(4)}\Big](t_r-\tau)\nonumber\\
U_{ij}^{{\rm 3.5PN}\,I_2I_4}(t_r)&=&-\frac{1}{432}M_{ab}M_{ijab}^{(7)}+\frac{1}{432}M_{ab}^{(1)}M_{ijab}^{(6)}-\frac{5}{756}M_{ab}^{(2)}M_{ijab}^{(5)}+\frac{19}{648}M_{ab}^{(3)}M_{ijab}^{(4)}+\frac{1957}{3024}M_{ab}^{(4)}M_{ijab}^{(3)}\nonumber\\
&+&\frac{1685}{1008}M_{ab}^{(5)}M_{ijab}^{(2)}+\frac{41}{28}M_{ab}^{(6)}M_{ijab}^{(1)}+\frac{91}{216}M_{ab}^{(7)}M_{ijab}\,,\nonumber\\
U_{ij}^{{\rm 3.5PN}\,I_3 I_3}(t_r)&=&-\frac{5}{252}M_{ab\langle i}M_{j\rangle ab}^{(7)}+\frac{5}{189}M_{ab\langle i}^{(1)}M_{j\rangle ab}^{(6)}+\frac{5}{126}M_{ab\langle i}^{(2)}M_{j\rangle ab}^{(5)}+\frac{5}{2268}M_{ab\langle i}^{(3)}M_{j\rangle ab}^{(4)}\,,\nonumber\\
U_{ij}^{{\rm 3.5PN}\, J_1 J_3}(t_r)&=&U_{ij}^{{\rm 3.5PN}\, J_2J_2}(t_r)=U_{ij}^{{\rm 3.5PN} \, J_3I_2}(t_r)=U_{ij}^{{\rm 3.5PN}\, I_3J_2}(t_r)=0\nonumber\\
\eea

A remarkable fact of the case under consideration (radial fall) is that all magnetic moments vanish identically, $V_L=S_L=J_L=0$, and we will not explicitly show these contributions when specializing general formulas to the present situation. All the needed multipole moments (beyond the quadrupole term, which has been shown explicitly above) are listed in Table \ref{tab1:variousUV} below. For the general expression see Ref. \cite{Blanchet:2013haa}. 

\begin{table*}
\caption{\label{tab1:variousUV} Eletric-type radiative multipoles $U_L$ (expressed in terms of canonical multipoles $M_L$) entering the waveform up to 3.5PN. All  magnetic-type moments vanish identically for the case under consideration, $V_L=S_L=J_L=0$, and are not displayed.
Here $C_{I_3}=\frac{2b_0}{c}e^{-\frac{97}{60}}$, $C_{I_4}=\frac{2b_0}{c}e^{-\frac{59}{30}}$,  $C_{I_5}=\frac{2b_0}{c}e^{-\frac{232}{105}}$. Moreover, $\mathcal{M}=E/c^2$. However, as a consequence of the fact that at infinity the particle starts its motion at rest $\gamma=1=\sqrt{1+p_\infty^2}$, i.e.,  $p_\infty=0$,  where $p_\infty$ denotes the velocity-at-infinity, we have $\mathcal{M}=M$.}
\begin{ruledtabular}
\begin{tabular}{ll}
$U_{ijk}(t_r)$ &$ M^{(3)}_{ijk}+\eta^3U_{ijk}^{\rm 1.5PN}+\eta^5 U_{ijk}^{\rm 2.5PN}+\eta^6U_{ijk}^{\rm. 3PN}+O\left(\eta^7\right)$\\
$U_{ijk}^{\rm 1.5PN}(t_r)$& $2{\mathcal M}\int_0^\infty d\tau M^{(5)}_{ijk}(t_r{-}\tau)\ln\left( \frac{\tau}{C_{I_3}}\right)$\\
$U_{ijk}^{\rm 2.5PN}(t_r)$&$\int_0^\infty d\tau\Big[-\frac{1}{3}M_{a\langle i}^{(3)}M_{jk\rangle a}^{(4)}\Big](t_r-\tau)+\frac{1}{4}M_{a\langle i}M_{jk\rangle a}^{(6)}+\frac{1}{4}M_{a\langle i}^{(1)}M_{jk\rangle a}^{(5)}$\\
&$+\frac{1}{4}M_{a\langle i}^{(2)}M_{jk\rangle a}^{(4)}-\frac{4}{3}M_{a\langle i}^{(3)}M_{jk\rangle a}^{(3)}-\frac{9}{4}M_{a\langle i}^{(4)}M_{jk\rangle a}^{(2)}-\frac{3}{4}M_{a\langle i}^{(5)}M_{jk\rangle a}^{(1)}+\frac{1}{12}M_{a\langle i}^{(6)}M_{jk\rangle a}$\\
$U_{ijk}^{\rm 3PN}(t_r)$&$2M^2\int_0^\infty d\tau M_{ijk}^{(6)}(t_r-\tau)\Big[\log^2\left(\frac{\tau}{2b_0}\right)+\frac{97}{30}\log\left(\frac{\tau}{2b_0}\right)-\frac{13}{21}\log\left(\frac{\tau}{2r_0}\right)+\frac{13283}{8820}\Big]$\\
\hline
$U_{ijkl}(t_r)$&$M^{(4)}_{ijkl}+\eta^3 U_{ijkl}^{\rm 1.5PN}+\eta^5U_{ijkl}^{\rm 2.5PN}+O\left(\eta^6\right)$\\
$U_{ijkl}^{\rm 1.5PN}(t_r)$&$2\mathcal{M}\int_0^\infty d\tau M_{ijkl}^{(6)}(t_r{-}\tau)\ln\left( \frac{\tau}{C_{I_4}}\right) 
+\frac{2}{5}\int_0^\infty d\tau\left(M^{(3)}_{\langle ij}M^{(3)}_{kl\rangle}\right)(t_r-\tau)-\frac{21}{5}M_{\langle ij}M_{kl\rangle}^{(5)}$\\
&$-\frac{63}{5}M_{\langle ij}^{(1)}M_{kl\rangle}^{(4)}-\frac{102}{5}M_{\langle ij}^{(2)}M_{kl\rangle}^{(3)}$\\
$U_{ijkl}^{\rm 2.5 PN}(t_r)$&$\int_0^\infty d\tau\Big[\frac{12}{55}M_{a\langle i}^{(4)}M_{jkl\rangle a}^{(4)}{-}\frac{14}{99}M_{a\langle ij}^{(4)}M_{kl\rangle a}^{(4)}\Big](t_r{-}\tau)$\\
&$+\frac{7}{55}M_{a\langle i}M_{jkl\rangle a}^{(7)}+\frac{7}{55}M_{a\langle i}^{(1)}M_{jkl\rangle a}^{(6)}+\frac{1}{25}M_{a\langle i}^{(2)}M_{jkl\rangle a}^{(5)}-\frac{28}{11}M_{a\langle i}^{(3)}M_{jkl\rangle a}^{(4)}-\frac{273}{55}M_{a\langle i}^{(4)}M_{jkl\rangle a}^{(3)}$\\
&$-\frac{203}{55}M_{a\langle i}^{(5)}M_{jkl\rangle a}^{(2)}-\frac{49}{55}M_{a\langle i}^{(6)}M_{jkl\rangle a}^{(1)}+\frac{14}{275}M_{a\langle i}^{(7)}M_{jkl\rangle a}+\frac{14}{33}M_{a\langle ij}M_{kl\rangle a}^{(7)}+\frac{37}{33}M_{a\langle ij}M_{kl\rangle a}^{(6)}$\\
&$+\frac{9}{11}M_{a\langle ij}^{(2)}M_{kl\rangle a}^{(5)}+\frac{8}{33}M_{a\langle ij}^{(3)}M_{kl\rangle a}^{(4)}$\\
\hline
$U_{ijklm}(t_r)$&$M_{ijklm}^{(5)}+\eta^3U_{ijklm}^{\rm 1.5PN}+O\left(\eta^5\right)$\\
$U_{ijklm}^{\rm 1.5PN}(t_r)$&$2\mathcal{M}\int_0^\infty d\tau M_{ijklm}^{(7)}(t_r-\tau)\ln\left( \frac{\tau}{C_{I_5}}\right) 
+\frac{20}{21}\int_0^\infty d\tau\Big[M_{\langle ij}^{(3)}M_{klm\rangle}^{(4)}\Big](t_r-\tau)$\\
&$-\frac{15}{7}M_{\langle ij}M_{klm\rangle}^{(6)}-\frac{41}{7}M_{\langle ij}^{(1)}M_{klm\rangle }^{(5)}-\frac{120}{7}M_{\langle ij}^{(2)}M_{klm\rangle}^{(4)}-\frac{710}{21}M_{\langle ij}^{(3)}M_{klm\rangle}^{(3)}-\frac{265}{7}M_{\langle ij}^{(4)}M_{klm\rangle}^{(2)}$\\
&$-\frac{155}{7}M_{\langle ij}^{(5)}M_{klm\rangle}^{(1)}-\frac{34}{7}M_{\langle ij}^{(6)}M_{klm\rangle}$\\
$U_{ijklmn}(t_r)$& $M_{ijklmn}^{(6)}+O\left(\eta^2\right)$\\
\hline
$U_{ijklmnp}(t_r)$&$M_{ijklmnp}^{(7)}+O\left(\eta\right)$\\
\end{tabular}
\end{ruledtabular}
\end{table*}
\end{widetext}
and where $b_0$ is the scale entering in the definition of the retarded time, and $r_0$ is the scale entering in the Partie Finie prescription of the MPM formalism. 

To proceed further let us note that
1) due to the property $M_L=I_L+O(\eta^5)$ (see e.g., \eqref{blocks_0}) in all terms in Eq. \eqref{lista_up_2.5pn} one can replace $M_{ij}$ as $I_{ij}$;
2) as already stated above, in the radial motion case   $V_L=S_L=J_L=0$;
3) all nonvanishing eletric type multipoles can be written as proportional to a {\it constant} STF tensor; for example,  $I_{ij}=f_2(t)q_{ij}$ with 
$f_2(t)\propto \nu t^{4/3}+O(\eta^2)$ and
$q_{ij}={\rm diag}\left[1,-\frac12, -\frac12\right]$ a constant STF $3\times 3$ matrix, which will imply additional special simplifications, e.g.,
\beq
q_{\langle ij\rangle }^2= q_{ij}^2-\frac12 \delta_{ij}=\frac{1}{2}q_{ij}\,,
\eeq
and ${\rm Tr}(q_{ij}^2)=\frac32$, besides the obvious advantage of simplifying most of the tensorial relations, e.g., 
\beq
M^{(5)}_{a\langle i} M^{}_{j\rangle a}{=} f_2^{(5)}(t)f_2(t) q_{\langle ij\rangle }^2{=}\frac12 f_2^{(5)}(t)f_2(t)q_{ij} \,. 
\eeq

\section{Motion in the presence of a radiation-reaction acceleration at the 3.5PN accuracy}

The relative motion of the system with respect of the cm is described by the relative acceleration ${\mathbf a}={\mathbf a}_1-{\mathbf a}_2$ (with ${\mathbf a}_a$, $a=1,2$ the accelerations of the two bodies). Its expression is formally given by
\bea
{\mathbf a}={\mathbf a}_{\rm N}+{\mathbf a}_{\rm 1PN}+ {\mathbf a}_{\rm 2PN}+{\mathbf a}_{\rm 2.5PN}+{\mathbf a}_{\rm 3PN}+{\mathbf a}_{\rm 3.5PN}\,,
\eea 
where  ${\mathbf a}_{\rm N}$, ${\mathbf a}_{\rm 1PN}$, ${\mathbf a}_{\rm 2PN}$ and ${\mathbf a}_{\rm 3PN}$ can be found in 
Eqs. (358) of Ref. \cite{Blanchet:2013haa}. The 2.5PN acceleration in harmonic coordinates originally due to Damour and Deruelle \cite{Damour:1981bh},  while the harmonic coordinates 3.5PN acceleration has been derived by Nissanke and Blanchet \cite{Nissanke:2004er} about 20 years later [We note, in passing, that even the 4.5PN acceleration is now, i.e. about 20 years after Nissanke and Blanchet, fully known \cite{Blanchet:2026suq}.]
It reads (in harmonic coordinates)
\bea
{\bf a}_{\rm 3.5PN}&=&  -\frac{GM}{r^2}\epsilon \Big[\Big(A_{2.5} +A_{3.5} \eta^2\Big){\bf n}\nonumber\\
&+&\Big(B_{2.5} +B_{3.5}\eta^2\Big){\bf v}\Big]\,,
\eea
with 
\bea
A_{2.5}&=&-\frac{8GM\nu}{5r}\dot r \left(\frac{17GM}{3r}+3v^2\right)\,,\nonumber\\ 
B_{2.5}&=& \frac{8GM\nu}{5r}\left(\frac{3GM}{r}+v^2\right)\,,\nonumber\\
A_{3.5}&=&\frac{GM\nu}{r}\dot{r}\Big[\frac{G^2M^2}{r^2}\Big(\frac{3956}{35}+\frac{184}{5}\nu\Big)\nonumber\\
&+&\frac{GM\nu^2}{r}\Big(\frac{692}{35}-\frac{724}{15}\nu\Big)+v^4\Big(\frac{366}{35}+12\nu\Big)\nonumber\\
&+&\frac{GM \dot{r}^2}{r}\Big(\frac{294}{5}+\frac{376}{5}\nu\Big)-v^2\dot{r}^2(114+12\nu)\nonumber\\
&+&112\dot{r}^4\Big]\,,\nonumber\\
B_{3.5}&=&\frac{GM\nu}{r}\Big[\frac{G^2M^2}{r^2}\Big(\frac{-1060}{21}-\frac{104}{5}\nu\Big)\nonumber\\
&+&\frac{GM\nu^2}{r}\Big(\frac{164}{21}+\frac{148}{5}\nu\Big)+v^4\Big(\frac{-626}{35}-\frac{12}{5}\nu\Big)\nonumber\\
&+&\frac{GM\dot{r}^2}{r}\Big(\frac{-82}{3}-\frac{848}{15}\nu\Big)-120\dot{r}^4\nonumber\\
&+&v^2\dot{r}\Big(\frac{678}{5}+\frac{12}{5}\nu\Big)\Big]\,.
\eea
Noticeably, for the absolute 3.5PN accuracy (i.e., fractional 1PN accuracy) we will only need the conservative motion to be fully known at the fractional 1PN order, i.e., 
\bea
{\bf a}_{\rm cons}={\mathbf a}_{\rm N}+\eta^2 {\mathbf a}_{\rm 1PN}\,.
\eea
The (1-dimensional) equations of motion $\ddot x=a_{\rm cons}+\epsilon a_{\rm rr}$ are easily solved looking for solutions of the type $x(t)=x_{\rm cons}(t)+\epsilon x_{\rm rr}(t)$, namely
\bea
\ddot x_{\rm cons}(t)+\epsilon \ddot x_{\rm rr}(t)&=& a_{\rm cons}(x_{\rm cons},\dot x_{\rm cons})\nonumber\\
&+&\frac{\partial a_{\rm cons}}{\partial x}\big|_{\rm cons} \epsilon x_{\rm rr} + \frac{\partial a_{\rm cons}}{\partial \dot x} \big|_{\rm cons} \epsilon \dot x_{\rm rr}\nonumber\\ 
&+& \epsilon a_{\rm rr}(x_{\rm cons},\dot x_{\rm cons})+O(\epsilon^2)\,,
\eea
that is
\bea
\ddot x_{\rm rr}(t)&=& \frac{\partial a_{\rm cons}}{\partial x}\big|_{\rm cons}   x_{\rm rr} + \frac{\partial a_{\rm cons}}{\partial \dot x} \big|_{\rm cons}  \dot x_{\rm rr}\nonumber\\ 
&+&  a_{\rm rr}(x_{\rm cons},\dot x_{\rm cons})\,.
\eea
Let us note  that the same simplicity is lost in the case of 2-dimensional motion, in which case the $x$ and $y$ motions are always coupled, and one conveniently uses a different method to solve the equations of motion, e.g., the method of variation of the constants \cite{Damour:2004bz,Bini:2022enm,Bini:2025rng}.
Solving for the motion corresponding to \lq\lq release from rest at infinity" in Cartesian coordinates adapted to the cm system we find the following solution for $x(t)\equiv x_p(t)$
\bea
\label{orbit}
x_p(t)&=&\frac{3^{\frac{2}{3}}
    M^{\frac{1}{3}} t^{\frac{2}{3}}}{2^{\frac{1}{3}}}+\frac{5}{2} M\eta ^2 (\nu -2)  \nonumber\\
   &+&\eta ^4 \frac{\left(5 \nu^2-19\nu +48\right) 
   M^{\frac{5}{3}}}{2\cdot 6^{\frac{2}{3}}t^{\frac{2}{3}}}\nonumber\\
   &+&\eta ^5\epsilon \frac{64  \nu   M^2
   }{63 t}+\frac{M^{\frac{7}{3}}\eta^6}{7560\cdot 6^\frac{1}{3}t^\frac{4}{3}}\Big(-134400\nonumber\\
   &+&(197264+4305\pi^2)\nu-27300\nu^2+2940\nu^3\Big)\nonumber\\
   &+&\epsilon\eta^7\frac{8\cdot 2^\frac{1}{3}M^\frac{8}{3}\nu(111+224\nu)}{567\cdot 3^\frac{2}{3}t^\frac{5}{3}}+O\left(\eta^8\right)\,,\nonumber\\
y_p(t)&=&0\,,
\eea
where, we recall, $\eta=\frac{1}{c}$ is a PN place-holder and $\epsilon$ is a radiation-reaction place-holder.  The masses $m_1$ and $m_2$ ($m_1>m_2$) of the bodies form the standard combinations: $m_1+m_2=M$ (total mass), $\mu=m_1m_2/(m_1+m_2)$ (reduced mass), $q=m_2/m_1$ (mass ratio), $\nu=m_1m_2/(m_1+m_2)^2$ (symmetric mass ratio). The radial fall is assumed to happen along the $x$-axis.

The conservative part of the orbit was found in Ref. \cite{Bini:2026cpw} (using a different parametrization). In the present case, the particle starts its motion at $r=\infty$  at $t=\infty$, at rest, and then, as time decreases, it moves towards the horizon (equivalent to have chosen $\dot r>0$ while falling). The final approach to the horizon cannot   be followed within a PN analysis, since the PN expansion breaks down when the particle enters the strong-field regime. This argument, as well as the limitations of this kind of computation, are extensively discussed in \cite{Maggiore:2007ulw}. The radiation-reaction part has been obtained by imposing, as it is customary, that it vanishes at the beginning of the process, in this case $t\to +\infty$ (where the motion starts). 
Below, for convenience, we will introduce a dimensionless time parameter $T=t/M$, so that for example Eq. \eqref{orbit} becomes 
\bea
\label{orbitT}
\frac{x_p(t)}{M}&=&\frac{3^{\frac{2}{3}}
    T^{\frac{2}{3}}}{2^{\frac{1}{3}}}+\frac{5}{2}  \eta ^2 (\nu -2)  \nonumber\\
   &+&\eta ^4 \frac{\left(5 \nu^2-19\nu +48\right) 
   }{2\cdot 6^{\frac{2}{3}}T^{\frac{2}{3}}}\nonumber\\
   &+&\eta ^5\epsilon \frac{64  \nu   
   }{63 T}+\frac{ \eta^6}{7560\cdot 6^\frac{1}{3}T^\frac{4}{3}}\Big(-134400\nonumber\\
   &+&(197264+4305\pi^2)\nu-27300\nu^2+2940\nu^3\Big)\nonumber\\
   &+&\epsilon\eta^7\frac{8\cdot 2^\frac{1}{3} \nu(111+224\nu)}{567\cdot 3^\frac{2}{3}T^\frac{5}{3}}+O\left(\eta^8\right)\,,\nonumber\\
y_p(t)&=& 0\,.
\eea
Beware of the standard notation $n^i$ as the unit vector of the orbital radial direction, not to be confused with the 
first leg of the cm triad \eqref{spat_triad}, always clear from the context. 

\section{Expressing the radiative moments in terms of the source multipole moments}

When the two bodies motion is confined to the $\langle x,y\rangle$ plane
\beq
x^i=(x (t),y (t),0),\quad v^i=(\dot x(t),\dot y(t),0), 
\eeq
all the radiative moments are easily expressed in terms of the source multipole moments. 
for example, the (source) quadrupole mass multipole at 3.5PN accuracy is given by
\bea
\label{I_ij_rep}
I_{ij}&=&\nu M \Bigg\{\Bigg[A_1{-}\frac{24}{7}\nu\eta^5\frac{G^2M^2}{r^2}\dot rA_{\rm rr}\Bigg] x_{\langle i}x_{j\rangle}\nonumber\\
&{+}&\Bigg[A_2\eta^2\,r \dot r{+}\frac{48}{7}\nu \eta^5\frac{G^2M^2}{r}C_{\rm rr}\Bigg] x_{\langle i}v_{j\rangle}\nonumber\\
&+&\Bigg[A_3\eta^2\,r^2+\eta^5 M \nu r\, r' B_{\rm rr}\Bigg] v_{\langle i}v_{j\rangle}\Bigg\}\nonumber\\
&+& O\left(\eta^8\right)\,,
\eea
where $A_1$, $A_2$, $A_3$ can be found in Eqs. (3.20a,b,c)  of Ref. \cite{Mishra:2015bqa} while $A_{\rm rr}$, $B_{\rm rr}$ and $C_{\rm rr}$ (including $O(\eta^7)$ corrections) are listed in Eqs. $A_{3a,b,c}$ of Ref. \cite{Faye:2012we}.
In the present case, inserting the (further simplified)  orbit representation \eqref{orbit} in the expression \eqref{I_ij_rep} of $I_{ij}$ one finds
\bea
I_{ij}=f_2(t)q_{ij}\,,\qquad q_{ij}={\rm diag}\left[1,-\frac12, -\frac12\right]\,,
\eea
with
\bea\label{f2}
\frac{f_2(t)}{\nu M^3}&=&6^\frac{1}{3}  T^\frac{4}{3}+\eta^2 \frac{2^\frac{2}{3}T^\frac{2}{3}}{3^\frac{1}{3}}\left(-\frac{66}{7}+\frac{16\nu}{7}\right)\nonumber\\
&+&\eta^4\left(\frac{16 \nu ^2}{21}-\frac{28
   \nu }{3}+\frac{1327}{42}\right)\nonumber\\
   &+&\frac{272\cdot 2^{\frac{2}{3}}\eta^5\nu}{63\cdot 3^{\frac{1}{3}}T^{\frac{1}{3}}}\nonumber\\
   &+&\frac{2^{\frac{1}{3}}\eta^6}{3^{\frac{2}{3}}T^{\frac{2}{3}}}\Big[-\frac{376 \nu
   ^3}{6237}-\frac{1234 \nu
   ^2}{693}-\frac{41 \pi ^2 \nu
   }{36}\nonumber\\
   &{+}&\frac{22760 \nu
   }{693}{-}\frac{8922514}{155925}{+}\frac{856}{315}\log\Big(\frac{3^\frac{2}{3}MT^\frac{2}{3}}{2^\frac{1}{3}b_0}\Big)\Big]\nonumber\\
   &+&\frac{\eta^7}{T}\Big(-\frac{1424\nu}{81}+\frac{11840\nu^2}{1701}\Big)+O\left(\eta^8\right)\,,
\eea 
where $T=\frac{t}{M}$ (dimensionless) and we took into account the big simplifications due to the diagonal character of $I_{ij}$.
For example, Eq. \eqref{UVdefs} implies
\bea
U_2&=&\nu M (c_\theta
    c_\phi+i s_\phi
   )^2\nonumber\\
   &\times&\Bigg\{\frac{1}{6^{\frac{2}{3}} T^{\frac{2}{3}}}+\frac{\eta^2}{6^{\frac{1}{3}}T^{\frac{4}{3}}}\left(\frac{11}{7}-\frac{8 \nu }{21}\right)\nonumber\\
   &+&\frac{\eta^3}{6^{\frac{2}{3}}T^{\frac{5}{3}}}\left(\frac{4 \mathcal{L}_{1,b_0}}{3}+\frac{2
   \pi }{3 \sqrt{3}}+\frac{1}{9}\right)\nonumber\\
   &+&\frac{2^{\frac{2}{3}}\eta^5}{3^{\frac{1}{3}}T^{\frac{7}{3}}}\Bigg[\frac{16 \pi  \nu }{63
   \sqrt{3}}-\frac{1658 \nu
   }{1323}-\frac{22 \pi }{21
   \sqrt{3}}+\frac{44}{9}\nonumber\\
   &+&\mathcal{L}_{1,b_0}\Big(\frac{44}{21}-\frac{32 \nu }{63}\Big)\Bigg]+\frac{2^\frac{1}{3}\eta^6}{3^\frac{2}{3}T^\frac{8}{3}}\Bigg[-\frac{13681408}{654885}\nonumber\\
   &-&\frac{470 \nu
   ^3}{18711}-\frac{3085 \nu
   ^2}{4158}-\frac{205 \pi ^2 \nu
   }{432}+\frac{28450 \nu
   }{2079}\nonumber\\
   &+&\frac{5 \pi
   ^2}{18}+\frac{167 \pi }{126
   \sqrt{3}}+\frac{10 \psi
   ^{(1)}\left(\frac{8}{3}\right)}
   {9}+\frac{10}{9}
   \mathcal{L}_{1,b_0}^2\nonumber\\
   &+&\left(\frac{41}{27}+\frac{10 \pi }{9
   \sqrt{3}}\right)
   \mathcal{L}_{1,b_0}+\frac{214
   \mathcal{L}_2}{189}\Bigg]\nonumber\\
   &+&\frac{\eta^7}{T^3}\Big(-\frac{58244\nu}{15309}+\frac{380\nu^2}{567}\Big)\Bigg\}\nonumber\\
   &+&O\left(\eta^8\right)\,.
   \eea
Similarly, all the other radiative moments up to $l=8$ have the following expressions
   \bea
U_3&=&\nu M \Delta c_\phi s_\theta (c_\theta c_\phi+i \,s_\phi)^2\nonumber\\
&\times&\Bigg\{\frac{\eta^2}{6^{2/3}T^{5/3}}\left(\frac{20 \nu }{27}-\frac{85}{27}\right)\nonumber\\
&+&\frac{\eta^4}{6^{1/3}T^{7/3}}\left(-\frac{20 \nu ^2}{81}+\frac{214
   \nu }{81}-\frac{211}{27}\right)\nonumber\\
   &+&\frac{\eta^5}{6^\frac{2}{3}T^\frac{8}{3}}\Bigg[\frac{100 \pi  \nu }{81
   \sqrt{3}}-\frac{1468 \nu
   }{1701}-\frac{425 \pi }{81
   \sqrt{3}}+\frac{85}{486}\nonumber\\
   &+&\left(\frac{200 \nu
   }{81}-\frac{850}{81}\right)
   \mathcal{L}_{1,b_0}\Bigg]\Bigg\}+O\left(\eta^7\right)\,,\nonumber\\
U_4&{=}&\nu M(7c_{2\theta}{-}3{-}14c_{2\phi}s_\theta^2)(c_\theta c_\phi{+}i\, s_\phi)^2\nonumber\\
&\times&\Bigg\{\frac{1}{6^{1/3}T^{4/3}}\left(\frac{5}{504}-\frac{5 \nu }{168}\right)\nonumber\\
&{+}&\frac{\eta^3}{6^{1/3}T^{7/3}}\Bigg[\left(\frac{5}{189}-\frac{5 \nu
   }{63}\right) \mathcal{L}_{1,b_0}\nonumber\\
   &+&\frac{5 \pi  \nu }{126
   \sqrt{3}}-\frac{89 \nu
   }{882}-\frac{5 \pi }{378
   \sqrt{3}}+\frac{11}{324}\Bigg]\nonumber\\
   &+&\frac{\eta^4}{6^\frac{2}{3}T^\frac{8}{3}}\Bigg(\frac{5 \nu ^3}{77}{-}\frac{475 \nu
   ^2}{693}{+}\frac{68899 \nu
   }{33264}{-}\frac{7907}{12936}\Bigg)\nonumber\\
 &+&\frac{\eta^5}{T^3}\Bigg(\frac{199 \nu
   ^2}{8019}-\frac{3169 \nu
   }{112266}\Bigg)\Bigg\}+O\left(\eta^6\right)\,,\nonumber\\
   \eea
   \bea
U_5&{=}&\nu M \Delta(1{+}3c_{2\theta}{-}6c_{2\phi}s_\theta^2)(c_\theta s_\phi{+}i s_\phi)^2c_\phi s_\theta\nonumber\\
&\times&\Bigg\{\frac{1}{6^\frac{2}{3}T^\frac{5}{3}}\Bigg(\frac{7 \nu }{108}-\frac{7}{216}\Bigg)\nonumber\\
&+&\frac{\eta^2}{6^\frac{1}{3}T^\frac{7}{3}}\Big(-\frac{25 \nu ^2}{162}+\frac{125
   \nu }{324}-\frac{25}{162}\Big)\nonumber\\
   &+&\frac{2^\frac{1}{3}\eta^3}{3^\frac{2}{3}T^\frac{8}{3}}\Bigg[\frac{35 \pi  \nu }{648
   \sqrt{3}}-\frac{5641 \nu
   }{27216}-\frac{35 \pi }{1296
   \sqrt{3}}+\frac{8}{243}\nonumber\\
   &+&\left(\frac{35 \nu
   }{324}-\frac{35}{648}\right)
   \mathcal{L}_{1,b_0}\Bigg]\Bigg\}+O\left(\eta^5\right)\,,\nonumber\\
U_6&=&\nu M\left(c_{\theta } c_{\phi }+i
   s_{\phi }\right){}^2\nonumber\\
   &\times&\left(108 c_{2 \theta }{-}99 c_{4
   \theta }{-}264 c_{4 \phi }
   s_{\theta }^4\right.\nonumber\\
   &{+}&\left.48 \left(11 c_{2
   \theta }{+}1\right) c_{2 \phi }
   s_{\theta }^2{-}73\right)\nonumber\\
   &\times&\frac{\eta^2}{6^\frac{2}{3}T^\frac{8}{3}}\Bigg(-\frac{875 \nu
   ^2}{228096}+\frac{875 \nu
   }{76032}\nonumber\\
   &+&\frac{175}{114048 \nu
   }-\frac{175}{20736}\Bigg)+O\left(\eta^4\right)
   \nonumber\\
U_7&=&\nu M \Delta(\nu-1)(3\nu-1)c_\phi s_\theta(c_\theta c_\phi+i\, s_\phi)^2\nonumber\\
&\times&\frac{1}{67392\cdot 6^{1/3}T^{7/3}}\Bigg[487+44c_{2\theta}+429 c_{4\theta}\nonumber\\
&-&176(7+13c_{2\theta})c_{2\phi}s_\theta^2+1144c_{4\phi}s_\theta^4\Bigg]\nonumber\\
&+&O\left(\eta^3\right)\,,\nonumber\\
\eea
\bea
U_8&=&\nu M (c_{\theta } c_{\phi }+is_{\phi })^2\Big(1221 c_{2 \theta }-858 c_{4
   \theta }\nonumber\\
   &{+}&715 c_{6 \theta
   }{-}2288 c_{6 \phi } s_{\theta
   }^6{+}2288 \left(3 c_{2 \theta
   }{+}1\right) c_{4 \phi }
   s_{\theta }^4\nonumber\\
   &-&22 \left(52 c_{2
   \theta }+195 c_{4 \theta
   }+137\right) c_{2 \phi }
   s_{\theta }^2-566\Big)\nonumber\\
   &{\times}&\frac{5}{1824768}\Big(7 \nu ^3{-}14 \nu ^2{+}7 \nu {-}1\Big){+}O\left(\eta^2\right)\,,\nonumber\\
U_9&=&O\left(\eta\right)\,,
\eea
with
\bea
\mathcal{L}_{1,b_0}&=&\log\left(\frac{2b_0}{3\sqrt{3}ct}\right)\,,\nonumber\\
\mathcal{L}_{2}&=&\log\left(\frac{2^\frac{2}{3}M^\frac{1}{3}r_0}{3^\frac{5}{6}b_0t^\frac{1}{3}}\right)\,,\nonumber\\
\Delta&=&\sqrt{1-4\nu}=\frac{m_1-m_2}{M}\,.
\eea

\section{Radiative losses}

The knowledge of the radiative moments allows one to compute  the radiative losses of energy, angular momentum, linear momentum (and cm position). The following (angle-integrated) time-domain fluxes (with   corresponding 
frequency-domain  losses) hold
\bea
\frac{dE^{\rm rad}}{dt}&=&{\mathcal F}_E(t_r) \,,\nonumber\\
\frac{dJ^{\rm rad}_i}{dt}&=&{\mathcal F}_{J_i}(t_r) \,,\nonumber\\
\frac{dP^{\rm rad}_i}{dt}&=&{\mathcal F}_{P_i}(t_r) \,,
\eea
with
\bea
{\mathcal F}_E(t_r) &=& \frac{G}{c^5}\left\{
\frac15 U^{(1)}_{ij}U^{(1)}_{ij}\right.\nonumber\\
&+& \eta^2 \left[\frac{1}{189}U^{(1)}_{ijk}U^{(1)}_{ijk}+\frac{16}{45}V^{(1)}_{ij}V^{(1)}_{ij}\right]\nonumber\\
&+& \eta^4 \left[ \frac{1}{9072}U^{(1)}_{ijkm}U^{(1)}_{ijkm}+\frac{1}{84}V^{(1)}_{ijk}V^{(1)}_{ijk} \right]\nonumber\\
&+&\eta^6 \left[\frac{1}{594000}U^{(1)}_{ijkml}U^{(1)}_{ijkml}+\frac{4}{14175}  V^{(1)}_{ijkl}V^{(1)}_{ijkl}  \right]\nonumber\\
&+& \left. O(\eta^8)\right\}\,,
\eea
\bea
{\mathcal F}_{J_i}(t_r) &=& \frac{G}{c^5}\epsilon_{iab}\left\{
\frac25 U_{aj}U^{(1)}_{bj}\right.\nonumber\\
&+& \eta^2 \left[\frac{1}{63}U_{ajk}U^{(1)}_{bjk}+\frac{32}{45}V_{aj}V^{(1)}_{bj}\right]\nonumber\\
&+& \eta^4 \left[ \frac{1}{2268}U_{ajkl}U^{(1)}_{bjkl}+\frac{1}{28}V_{ajk}V^{(1)}_{bjk} \right]\nonumber\\
&+&\eta^6 \left[\frac{1}{118800}U_{ajklm}U^{(1)}_{bjklm}+\frac{16}{14175}  V_{ajkl}V^{(1)}_{bjkl}  \right]\nonumber\\
&+& \left. O(\eta^8)\right\}\,,
\eea
\bea
&&{\mathcal F}_{P_i}(t_r) = \frac{G}{c^7}\left\{
\frac2{63} U^{(1)}_{ijk}U^{(1)}_{jk}\right.\nonumber\\
&&+ \eta^2 \left[\frac{1}{1134}U^{(1)}_{ijkl}U^{(1)}_{jkl}+\frac{1}{126}\epsilon_{ijk}U^{(1)}_{jab}V^{(1)}_{kab}+\frac{4}{63}V^{(1)}_{ijk}V^{(1)}_{jk}\right]\nonumber\\
&&+ \eta^4 \left[ \frac{1}{59400}U^{(1)}_{ijklm}U^{(1)}_{jklm}+\frac{2}{14175}\epsilon_{ijk}U^{(1)}_{jabc}V^{(1)}_{kabc}\right.\nonumber\\
&&+\left.\left. \frac{2}{945}V^{(1)}_{ijkl}V^{(1)}_{jkl} \right]+ O(\eta^6)\right\}\,.
\eea
Concerning the flux associated with the cm position see  Sec. 2.4.1, Eq. 181b, of Ref. \cite{Blanchet:2013haa} 
and related discussion (including references therein). 

In the present case, since all the magnetic-type multipoles are vanishing, the above relations simplify as
\bea
\label{E_loss}
{\mathcal F}_E(t_r) &=& \frac{G}{c^5}\left\{
\frac15 U^{(1)}_{ij}U^{(1)}_{ij}+\eta^2  \frac{1}{189}U^{(1)}_{ijk}U^{(1)}_{ijk}\right.\nonumber\\
&+& \eta^4  \frac{1}{9072}U^{(1)}_{ijkm}U^{(1)}_{ijkm} \nonumber\\
&+&\eta^6  \frac{1}{594000}U^{(1)}_{ijkml}U^{(1)}_{ijkml} \nonumber\\
&+& \left. O(\eta^8)\right\}\,,
\eea
\bea
\label{J_loss}
{\mathcal F}_{J_i}(t_r) &=& \frac{G}{c^5}\epsilon_{iab}\left\{
\frac25 U_{aj}U^{(1)}_{bj}+\eta^2  \frac{1}{63}U_{ajk}U^{(1)}_{bjk}\right.\nonumber\\
&+& \eta^4  \frac{1}{2268}U_{ajkl}U^{(1)}_{bjkl} \nonumber\\
&+&\eta^6  \frac{1}{118800}U_{ajklm}U^{(1)}_{bjklm} \nonumber\\
&+& \left. O(\eta^8)\right\}\,,
\eea
\bea
\label{P_loss}
&&{\mathcal F}_{P_i}(t_r) = \frac{G}{c^7}\left\{
\frac2{63} U^{(1)}_{ijk}U^{(1)}_{jk}\right.\nonumber\\
&&+ \eta^2  \frac{1}{1134}U^{(1)}_{ijkl}U^{(1)}_{jkl}  \nonumber\\
&&\left. + \eta^4  \frac{1}{59400}U^{(1)}_{ijklm}U^{(1)}_{jklm} + O(\eta^6)\right\}\,.
\eea
Consequently
\bea
\Delta E_{\rm rad}&=&\int dt_r {\mathcal F}_E(t_r)\,,\nonumber\\
\Delta J^i_{\rm rad}&=&\int dt_r {\mathcal F}_{J_i}(t_r)\,,\nonumber\\
\Delta P^i_{\rm rad}&=&\int dt_r {\mathcal F}_{P_i}(t_r)\,.
\eea

As an application of our results let us first compute the energy loss of the system.
The latter can be obtained directly from the waveform according to the relation
\beq
\frac{dE_{\rm rad}}{dt}=\frac{1}{16\pi}\int d\Omega |\dot h|^2,\qquad \dot h=4\dot {\mathcal W}\,,
\eeq
and can be computed multipole-by-multipole (i.e., isolating the contribution of each $l$ separately).
For example, the contribution of the mass quadrupole to the energy loss is
\begin{widetext}
\bea\label{dedteta}
\frac{1}{\nu^2}\left.\frac{dE_{\rm rad}}{dt}\right|_{2}&=&\frac{32\cdot 2^\frac{2}{3}}{405
   \cdot3^\frac{1}{3} T^\frac{10}{3}}+\frac{\eta^2}{T^4}\left(\frac{2816}{2835}-\frac{2048 \nu
   }{8505}\right)+\frac{2^\frac{2}{3}\eta^3}{3^\frac{1}{3}T^\frac{13}{3}}\left(\frac{1312}{3645}+\frac{64 \pi
   }{243 \sqrt{3}}+\frac{128 \mathcal{L}_{1,b_0}}{243}\right)\nonumber\\
   &{+}&\frac{2^\frac{1}{3}\eta^4}{3^\frac{2}{3}T^\frac{14}{3}}\left(\frac{16384 \nu
   ^2}{59535}{-}\frac{45056 \nu
   }{19845}{+}\frac{30976}{6615}\right){+}\frac{\eta^5}{T^5}\Big[\frac{4096 \pi  \nu }{25515
   \sqrt{3}}{-}\frac{97024 \nu
   }{25515}{-}\frac{5632 \pi
   }{8505
   \sqrt{3}}{+}\frac{18304}{1215}\nonumber\\
   &+&\Big(\frac{22528}{2835}-\frac{16384
   \nu }{8505}\Big)\mathcal{L}_{1,r_0}\Big]+\frac{2^\frac{2}{3}\eta^6}{3^\frac{1}{3}T^\frac{16}{3}}\Bigg[-\frac{48128 \nu
   ^3}{1515591}-\frac{157952 \nu
   ^2}{168399}-\frac{1312 \pi ^2
   \nu }{2187}+\frac{2913280 \nu
   }{168399}\nonumber\\
   &+&\frac{928 \pi
   ^2}{2187}+\frac{214496 \pi
   }{76545
   \sqrt{3}}-\frac{6657761816}{265
   228425}+\frac{1664}{729}
   \mathcal{L}_{1,b_0}^2+\left(\frac{45632}{10935}+\frac{1664 \pi
   }{729 \sqrt{3}}\right)
   \mathcal{L}_{1,b_0}+\frac{10956
   8
   \mathcal{L}_2}{76545}\nonumber\\
   &+&\frac{102
   4 \psi
   ^{(1)}\left(\frac{8}{3}\right)}
   {729}\Bigg]+\frac{2^\frac{1}{3}\eta^7}{3^\frac{2}{3}T^\frac{17}{3}}\Bigg[-\frac{32768 \pi  \nu ^2}{25515
   \sqrt{3}}+\frac{788224 \nu
   ^2}{76545}+\frac{90112 \pi  \nu
   }{8505 \sqrt{3}}-\frac{24490240
   \nu }{321489}\nonumber\\
   &-&\frac{61952 \pi
   }{2835
   \sqrt{3}}+\frac{7186432}{59535}+\left(\frac{65536 \nu
   ^2}{25515}-\frac{180224 \nu
   }{8505}+\frac{123904}{2835}\right) \mathcal{L}_{1,b_0}\Bigg]\nonumber\\
   &+&O\left(\eta^8\right)\,.
   \eea

Including all multipolar contributions (up to $l=9$), one gets the final time-domain 3.5PN accurate result 
\bea
\label{DEDT}
\frac{1}{\nu^2}\frac{dE_{\rm rad}}{dt}&=&\frac{32\cdot 2^\frac{2}{3}}{405
   \cdot 3^\frac{1}{3} T^\frac{10}{3}}{+}\frac{\eta^2}{T^4}\left(\frac{2816}{2835 }{-}\frac{2048 \nu }{8505 }\right) {+}\frac{2^\frac{2}{3}\eta^3}{3^\frac{1}{3}T^\frac{13}{3}}\left(\frac{1312}{3645}{+}\frac{64 \pi }{243\sqrt{3}}+\frac{128
   \mathcal{L}_{1,b_0}}{243}\right) \nonumber\\
   &+&\frac{2^\frac{1}{3}\eta^4}{3^\frac{2}{3}T^\frac{14}{3}}\left(\frac{1076224
   }{229635}-\frac{25088 \nu
   }{10935}+\frac{7936 \nu^2}{25515}\right)+\frac{\eta^5}{T^5}\Big(\frac{18304}{1215}-\frac{5632 \pi }{8505 \sqrt{3}}-\frac{97024 \nu }{25515}\nonumber\\
   &+&\frac{4096 \pi  \nu }{25515
   \sqrt{3}}+\Big(\frac{22528}{2835}-\frac{16384 \nu }{8505}\Big)\mathcal{L}_{1,b_0}\Big)+\frac{2^\frac{2}{3}\eta^6}{3^\frac{1}{3}T^\frac{16}{3}}\Bigg[-\frac{293248 \nu
   ^3}{1515591}+\frac{71648 \nu
   ^2}{168399}-\frac{1312 \pi ^2
   \nu }{2187}\nonumber\\
   &+&\frac{21476960 \nu
   }{1515591}+\frac{928 \pi
   ^2}{2187}+\frac{214496 \pi
   }{76545
   \sqrt{3}}-\frac{6472205816}{265
   228425}+\frac{1664}{729}
   \mathcal{L}_{1,b_0}^2+\left(\frac{45632}{10935}+\frac{1664 \pi
   }{729 \sqrt{3}}\right)
   \mathcal{L}_{1,b_0}\nonumber\\
   &+&\frac{10956
   8
   \mathcal{L}_2}{76545}+\frac{102
   4 \psi
   ^{(1)}\left(\frac{8}{3}\right)}
   {729}\Bigg]+\frac{2^\frac{1}{3}\eta^7}{3^\frac{2}{3}T^\frac{17}{3}}\Bigg[-\frac{15872 \pi  \nu ^2}{10935
   \sqrt{3}}+\frac{92416 \nu
   ^2}{8505}+\frac{351232 \pi  \nu
   }{32805
   \sqrt{3}}-\frac{10548224 \nu
   }{137781}\nonumber\\
   &-&\frac{2152448 \pi
   }{98415
   \sqrt{3}}+\frac{249603328}{2066
   715}+\left(\frac{31744 \nu
   ^2}{10935}-\frac{702464 \nu
   }{32805}+\frac{4304896}{98415}\right) \mathcal{L}_{1,b_0}\Bigg]+O\left(\eta^8\right)\,.
\eea
\begin{figure}
\includegraphics[scale=0.85]{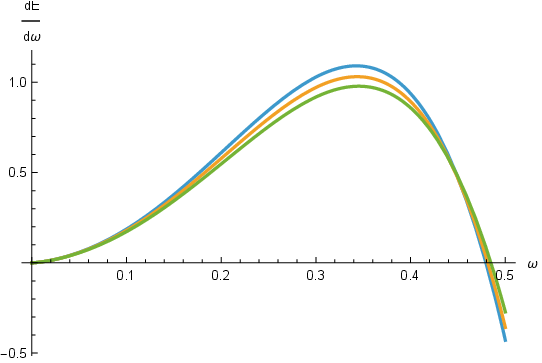}
\includegraphics[scale=0.85]{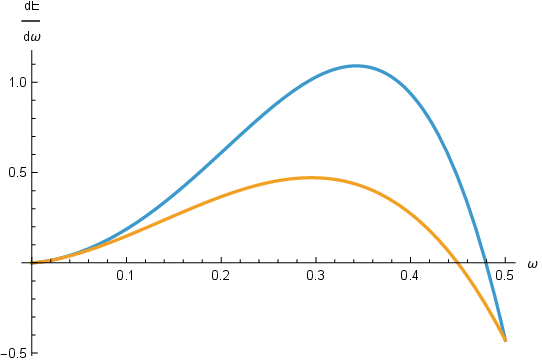}
\caption{\label{fig:1} Plots of the energy flux in the frequency domain, \eqref{dedomega}.
Left panel: comparison of $dE/d\omega$ for $\nu=0,1/8,1/4$, shown by the blue online, orange online, and green online curves, respectively.
Right panel: comparison of $dE/d\omega$ at 2.5 PN (orange online curve) and 3.5 PN (blue online curve) for $\nu=0$.
In both plots we used $M=\eta=1$ and $r_0=b_0$. The maximum of the energy loss moves from $M\omega=0.295142$ (2.5 PN) \cite{Bini:2026ova} to $M\omega=0.342657$ (3.5 PN), which is closer to the maximum obtained from numerical integration in \cite{Detweiler:1979xr}, namely $M\omega=0.36$ (see Fig. 2 of the aforementioned paper).}
\end{figure}
\begin{figure}
\includegraphics[scale=0.85]{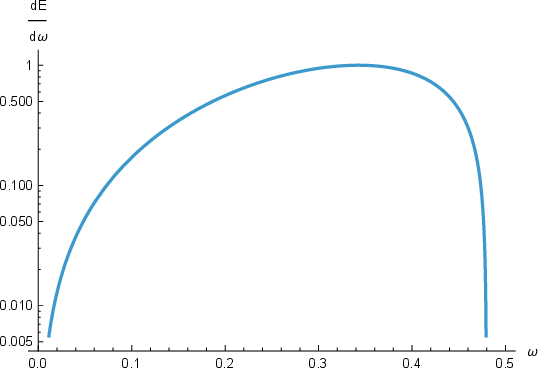}
\includegraphics[scale=0.85]{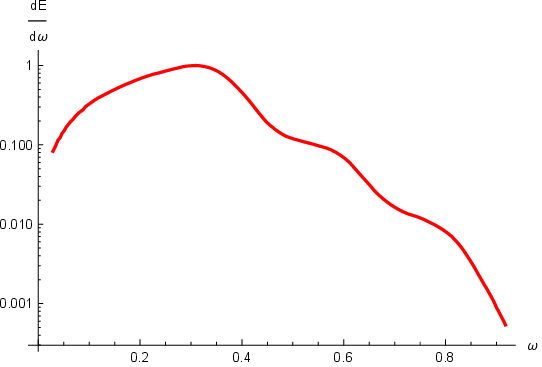}
\caption{\label{fig:2} We show a comparison between the energy flux in the frequency domain at 3.5 PN (blue online curve) Eq. \eqref{dedomega}, and the same quantity shown in Fig. 2 of \cite{Detweiler:1979xr} (red online curve). In that work, the radial infall corresponds to the lowest curve $J/(\mu M)=0$, which we have extracted from the original paper by using the WebPlotDigitizer application \cite{webplotdigitizer}. The points extracted though the digitalization are available in the ancillary file of this paper \cite{anc}.}
\end{figure}
\begin{figure}
\includegraphics[scale=0.85]{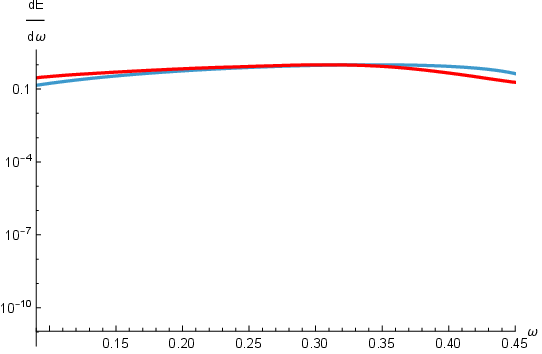}
\caption{\label{fig:3} Detewiler's results \cite{Detweiler:1979xr} (extracted from the original paper by using  WebPlotDigitizer \cite{webplotdigitizer}) and our actual results are superposed showing nice agreement in a region of the frequency domain which is safe from both numerical problems (i.e., avoiding very small frequencies where high numerical precision is needed) and analytical problems (avoiding very high frequencies, where the PN approximation is not expected to be valid anymore).}
\end{figure}
In the limit $\nu=0$, the right-hand-side of Eq. \eqref{DEDT} coincides with the expansion appearing in Eq. 5.36 of \cite{Bini:2026cpw}. Let us also note that Eq. \eqref{DEDT} represents an energy flux (rather than the energy loss itself, which instead arises only after integration). Therefore, there is no issue with the dependence on the regularization scale appearing in Eq. \eqref{DEDT}.

The total emitted energy follows by integrating the PN-expanded result, Eq. \eqref{DEDT}, over the PN-validity time interval $t\in (t_{\rm max}, +\infty)$ (with our choice of parametrization of the orbit), as explained above.
Similarly, in the frequency domain we have formally 
\beq
\frac{dE_{\rm rad}}{d\omega}=\sum_{k=4}^\infty C_k (M\omega)^{k/3}\,,
\eeq
(with $C_5=0$),  or, explicitly
\bea\label{dedomega}
\frac{1}{\frac{2}{3}\pi^2\nu^2M^2}\frac{dE_{\rm rad}}{d\omega}&{=}&\frac{12\cdot 6^\frac{2}{3}  \Gamma
   \left(\frac{4}{3}\right)^2}{5\pi }(M \omega)^{\frac{4}{3}}{+}\eta^2\Big(\frac{176\sqrt{3}}{35}{-}\frac{128\nu}{35\sqrt{3}}\Big) (M\omega)^2{-}\frac{24}{5}\eta^3 6^{\frac{2}{3}}  \Gamma
   \left(\frac{4}{3}\right)^2(M \omega)^{\frac{7}{3}}\nonumber\\
   &{+}&\eta^4\frac{2^{\frac{1}{3}} \Gamma\left(\frac{2}{3}\right)^2}{3^{\frac{2}{3}} \pi }\Big(\frac{248 \nu^2}{35}{-}\frac{784 \nu }{15}{+}\frac{33632}{315}\Big)(M\omega)^{\frac{8}{3}}  {+}\eta^5\Big(\frac{256 \pi  \nu }{35\sqrt{3}}{+}\frac{48 \nu}{49}{-}\frac{352 \sqrt{3} \pi}{35}\Big)(M\omega)^3\nonumber\\
   &+&\frac{2^\frac{2}{3}\Gamma\left(\frac{1}{3}\right)^2}{3^\frac{1}{3}\pi}\eta^6(M\omega)^\frac{10}{3}\Bigg[-\frac{256 \nu ^3}{495}+\frac{2736
   \nu ^2}{385}+\frac{41 \pi ^2
   \nu }{30}-\frac{528244 \nu
   }{10395}+\frac{16 \pi
   ^2}{15}+\frac{120960797}{181912
   5}\nonumber\\
   &-&\frac{1712}{525} \log
   \left(\frac{2^\frac{2}{3}
   e^{\frac{\gamma }{3}+\frac{\pi
   }{3 \sqrt{3}}} (M\omega)^\frac{1}{3}
   r_0 }{3^\frac{1}{3}
   b_0}\right)\Bigg]+\frac{2^\frac{1}{3}\eta^7}{3^\frac{2}{3}}(M\omega)^\frac{11}{3}\Bigg[\left(-\frac{496 \nu
   ^2}{35}+\frac{1568 \nu
   }{15}-\frac{67264}{315}\right)
   \Gamma
   \left(\frac{2}{3}\right)^2\nonumber\\
   &+&\left(\frac{465952 \nu
   }{8505}-\frac{608 \nu
   ^2}{63}\right) \Gamma
   \left(\frac{4}{3}\right)\Bigg]{+}O\left(\eta^8\right)\,,
\eea
\end{widetext}
where the relevant integrals appearing in the Fourier transforms computations are given by
\bea
\mathcal{J}_s(\omega)&=&\int_{-\infty}^\infty dt e^{i\omega t}t^s\nonumber\\
&=&2e^{\frac{i\pi s}{2}}|\omega|^{-1-s}\Gamma(1+s)(\Theta(\omega)-1)\sin(\pi s)\,,\nonumber\\
\eea
\bea
\mathcal{J}^{\rm log}_s(\omega)&=&\int_{-\infty}^\infty dt e^{i\omega t}t^s\log(t)=\frac{d\mathcal{J}_s(\omega)}{ds}\nonumber\\
&=&e^{\frac{i \pi  s}{2}} (\theta
   (\omega )-1) \Gamma (s+1) |
   \omega | ^{-s-1} \nonumber\\
   &\times&\left(i \sin
   (\pi  s) \left(\pi +i \log
   \left(\omega
   ^2\right)\right)+2 \pi  \cos
   (\pi  s)\right.\nonumber\\
   &+&\left.2 \sin (\pi  s) \psi
   ^{(0)}(s+1)\right)\,.
\eea
Plots of the the energy flux in the frequency domain \eqref{dedomega} are shown in Figs. \ref{fig:1} and \ref{fig:2}. 
Let us discuss now the remaining losses: angular momentum loss, linear momentum loss and cm motion.

The loss of angular momentum, Eq. \eqref{J_loss}
is identically zero in the present case, because of the radial character of the motion. 

The losses of linear momentum \eqref{P_loss} and the position of the cm are defined as
\bea\label{dPdG}
\frac{dP^i_{\rm rad}}{dt}&=&\eta^2\Big[\frac{2}{63}U^{(1)}_{ijk}U^{(1)}_{jk}{+}\frac{\eta^2}{1134}U^{(1)}_{ijkl}U^{(1)}_{jkl}\nonumber\\
&+&\frac{\eta^4}{59400}U^{(1)}_{ijklm}U^{(1)}_{jklm}\Big]+O\left(\eta^8\right)\,,\nonumber\\
\frac{dG^i_{\rm rad}}{dt}&=&\mathcal{F}^i_{\rm G}=\eta^2\Big[\frac{2}{21}U_{ijk}U^{(1)}_{jk}+\frac{2\eta^2}{567}U_{ijkl}U^{(1)}_{jkl}\nonumber\\
&+&\frac{\eta^4}{11880}U_{ijklm}U^{(1)}_{jklm}\Big]+O\left(\eta^8\right)\,,
\eea
having already considered the present simplification $V_{ij\dots}=0$. Plugging the relative orbit into these equations, we obtain
\bea
\frac{dP^1_{\rm rad}}{dt}&=& \nu ^2\Delta\eta^4\Bigg[\frac{2^\frac{2}{3}}{3^\frac{1}{3}T^\frac{13}{3}}\left(\frac{640 \nu
   }{15309}-\frac{2720}{15309}\right)\nonumber\\
   &+&\frac{\eta^2}{T^5}\Big(-\frac{512 \nu
   ^2}{6237}+\frac{100352 \nu
   }{120285}-\frac{17543872}{7577
   955}\Big)\nonumber\\
   &+&\frac{2^\frac{2}{3}\eta^3}{3^\frac{1}{3}T^\frac{16}{3}}\Big(\frac{8320 \pi  \nu }{45927
   \sqrt{3}}+\frac{486592 \nu
   }{4822335}-\frac{35360 \pi
   }{45927
   \sqrt{3}}\nonumber\\
   &-&\frac{102544}{137781}+\left(\frac{16640 \nu
   }{45927}-\frac{70720}{45927}\right) \mathcal{L}_{1,b_0}\Big)\Bigg]+O\left(\eta^8\right)\,,\nonumber\\
 \frac{dP^2_{\rm rad}}{dt}&=&\frac{dP^3_{\rm rad}}{dt}=0\,,
\eea
and
\bea
 \mathcal{F}_{\rm G}^1&=& \nu ^2 M\Delta \eta^4  \Bigg[\frac{2^{2/3}}{3^{1/3}
   T^{10/3}}\left(\frac{544}{1701}-\frac{128 \nu
   }{1701}\right)\nonumber\\
   &+&\frac{\eta^2}{T^4}\Big(\frac{40448 \nu
   ^2}{392931}-\frac{83456 \nu
   }{72765}+\frac{61677376}{17681
   895}\Big)\nonumber\\
   &+&\frac{2^\frac{2}{3}\eta^3}{3^\frac{1}{3}T^\frac{13}{3}}\Big(-\frac{1280 \pi  \nu }{5103
   \sqrt{3}}-\frac{44864 \nu
   }{535815}+\frac{5440 \pi
   }{5103
   \sqrt{3}}\nonumber\\
   &+&\frac{10880}{15309}+\left(\frac{10880}{5103}-\frac{25
   60 \nu }{5103}\right)
   \mathcal{L}_{1,b_0}\Bigg]+O\left(\eta^8\right)\,,\nonumber\\
  \mathcal{F}_{\rm G}^2&=&\mathcal{F}_{\rm G}^3=0\,,
\eea
coherently with the radial motion assumption (with the $x$-axis, i.e., $e_1$, as the axis of the motion). Remember that to restore the correct $\eta$  dependence, the energy flux \eqref{DEDT} and the other fluxes in \eqref{dPdG} should be multiplied by an overall $\eta^5$.

\subsection{Schott terms}

As we have seen above, within a Hamiltonian framework, the system (binding) energy and the angular momentum  are given by
\bea
E_{\rm sys}&=& H(x,p)-M c^2\,,\nonumber\\
J_{\rm sys}&=& p_\phi=0\,,
\eea 
where $E_{\rm sys}$ is the total energy of the system after removal of the rest energy of the particles. As already stated, the presence of the nonconservative force causes radiative losses: energy, angular momentum and linear momentum losses, plus cm motion. Therefore, the dynamics is regulated by the following  equations
\bea
\label{sys_dynamics}
\frac{dE_{\rm sys}}{dt}&=&\dot x^i \frac{\partial H(x,p)}{\partial x^i}+\dot p_i \frac{\partial H(x,p)}{\partial p_i}\nonumber\\
&=& \dot x^i {\mathcal F}_i=\frac{\partial H(x,p)}{\partial p_i}{\mathcal F}_i\,, \nonumber\\
\frac{dJ_{\rm sys}}{dt}&=& -\frac{\partial H(x,p)}{\partial \phi}+ {\mathcal F}_\phi=  {\mathcal F}_\phi\,,
\eea
where we have  taken into account that in our case $\frac{\partial H}{\partial \phi}=0$.
Notably, $\dot x^i=v^i\not= p^i/\mu$ but follows from Hamilton's equations $\dot x^i =\frac{\partial H}{\partial p_i}$. In general
$v^i=p^i/\mu+O(\eta^2)$.

Denoting as ${\mathcal F}_E$ and ${\mathcal F}_J$ the radiated energy and angular momentum fluxes at infinity 
the following balance laws hold
\bea
\label{E_J_losses}
\frac{dE_{\rm sys}}{dt}+\frac{dE_{\rm Schott}}{dt}&=& -{\mathcal F}_E=-\frac{dE_{\rm rad}}{dt}\,,\nonumber\\
\frac{dJ_{\rm sys}}{dt}+\frac{dJ_{\rm Schott}}{dt}&=& -{\mathcal F}_J=-\frac{dJ^i_{\rm rad}}{dt}\,.
\eea
The energy and angular momentum Schott terms represent  the (instantaneous) energy and angular momentum of the gravitational field.
The integrated version of Eqs. \eqref{E_J_losses} gives the mentioned losses
\bea
E_{\rm rad}&=& \int dt {\mathcal F}_E\,,\nonumber\\
J_{\rm rad}&=& \int dt {\mathcal F}_J\,,
\eea
but in general they do not coincide with \lq\lq kinematical terms," namely
\bea
E_{\rm rad}&\not =&-\int dt \frac{\partial H(x,p)}{\partial p_i}{\mathcal F}_i\,,\nonumber\\
J_{\rm rad}&\not =& -\int dt {\mathcal F}_\phi\,,
\eea
unless  the Schott terms decay  fast both at the horizon and at infinity. 
By using Eqs. \eqref{E_J_losses} we have computed $\frac{dE_{\rm Schott}}{dt}$ and $\frac{dJ_{\rm Schott}}{dt}$, 
\begin{widetext}
\bea
\frac{dE_{\rm Schott}}{dt}&=&\frac{2^\frac{2}{3}}{3^\frac{1}{3}T^\frac{10}{3}}\Big(-\frac{128 \nu
   ^2}{81}-\frac{32}{405}\Big)+\frac{\eta^2}{T^4}\Big(\frac{4352 \nu ^3}{243}-\frac{320
   \nu ^2}{81}+\frac{2048 \nu
   }{8505}-\frac{2816}{2835}\Big)+\frac{2^\frac{2}{3}\eta^3}{3^\frac{1}{3}T^\frac{13}{3}}\Big(-\frac{128}{243}
   \mathcal{L}_{1,b_0}\nonumber\\
   &-&\frac{64
   \pi }{243
   \sqrt{3}}-\frac{1312}{3645}\Big)+\frac{2^\frac{1}{3}\eta^4}{3^\frac{2}{3}T^\frac{14}{3}}\Big(-\frac{7936 \nu
   ^2}{25515}+\frac{25088 \nu
   }{10935}-\frac{1076224}{229635
   }\Big)+\frac{\eta^5}{T^5}\Big(\left(\frac{16384 \nu
   }{8505}-\frac{22528}{2835}\right)
   \mathcal{L}_{1,b_0}\nonumber\\
   &-&\frac{4096
   \pi  \nu }{25515
   \sqrt{3}}+\frac{97024 \nu
   }{25515}+\frac{5632 \pi }{8505
   \sqrt{3}}-\frac{18304}{1215}\Big)+\frac{2^\frac{2}{3}\eta^6}{3^\frac{1}{3}T^\frac{16}{3}}\Big(-\frac{1664}{729}
   \mathcal{L}_{1,b_0}^2+\left(-\frac{45632}{10935}-\frac{1664
   \pi }{729 \sqrt{3}}\right)
   \mathcal{L}_{1,b_0}\nonumber\\
   &{+}&\frac{2932
   48 \nu
   ^3}{1515591}{-}\frac{71648 \nu
   ^2}{168399}{+}\frac{1312 \pi ^2
   \nu }{2187}{-}\frac{21476960 \nu
   }{1515591}{-}\frac{109568
   \mathcal{L}_2}{76545}{-}\frac{92
   8 \pi ^2}{2187}{-}\frac{214496
   \pi }{76545
   \sqrt{3}}{+}\frac{6472205816}{26
   5228425}\nonumber\\
   &-&\frac{1024 \psi
   ^{(1)}\left(\frac{8}{3}\right)
   }{729}\Big)+\frac{2^\frac{2}{3}\eta^7}{3^\frac{1}{3}T^\frac{16}{3}}\Big(-\frac{1664}{729}
   \mathcal{L}_{1,b_0}^2+\left(-\frac{45632}{10935}-\frac{1664
   \pi }{729 \sqrt{3}}\right)
   \mathcal{L}_{1,b_0}+\frac{2932
   48 \nu
   ^3}{1515591}-\frac{71648 \nu
   ^2}{168399}\nonumber\\
   &+&\frac{1312 \pi ^2
   \nu }{2187}-\frac{21476960 \nu
   }{1515591}-\frac{109568
   \mathcal{L}_2}{76545}-\frac{92
   8 \pi ^2}{2187}-\frac{214496
   \pi }{76545
   \sqrt{3}}+\frac{6472205816}{26
   5228425}-\frac{1024 \psi
   ^{(1)}\left(\frac{8}{3}\right)
   }{729}\Big)\nonumber\\
   &+&O\left(\eta^8\right)\,,\nonumber\\
\frac{dJ_{\rm Schott}}{dt}&=&0\,.
\eea
\end{widetext}

Having obtained in Eqs. \eqref{orbit} the radiation-reacted orbit, we can also evaluate the variation of the (relative) linear momentum of the system. Namely, from the equations of motion
\bea
\label{eq_dot_pi}
\dot p_i&=&-\frac{\partial H(x,p)}{\partial x^i}+{\mathcal F}_i\,,
\eea
we find
\bea
\Delta p_i &=& -\int_{-\infty}^\infty dt \frac{\partial H(x,p)}{\partial x^i}\bigg|_{
\begin{array}{l}
x^i=x^i_{\rm cons}+\delta^{\rm rr}x^i\\
p_i=p_{i\rm cons}+\delta^{\rm rr}p_i
\end{array}
}\nonumber\\
&+&\int_{-\infty}^\infty dt {\mathcal F}_i\bigg|_{\begin{array}{l}
x^i=x^i_{\rm cons}+\delta^{\rm rr}x^i\\
p_i=p_{i\rm cons}+\delta^{\rm rr}p_i
\end{array}}\nonumber\\
&=& \Delta p_i^{\rm cons}+\Delta p_i^{\rm rr}\,.
\eea
Here we do not show the complete variation but the  \lq\lq instantaneous" one, as corresponding to an integral taken from $-\infty$ up to the actual time $t$. 
We find $\Delta p_\phi(t)=0$ and
\bea
\Delta p_r(t) &=& \frac{2^\frac{2}{3}}{3^\frac{1}{3}T^\frac{1}{3}} +\frac{\eta^2}{T}\Big(\frac{4}{3}-\frac{14 \nu }{9}\Big)\nonumber\\
&+&\frac{2^\frac{1}{3}\eta^4}{3^\frac{2}{3}T^\frac{5}{3}}\Big(\frac{169 \nu ^2}{30}-\frac{187
   \nu }{30}-\frac{8}{5}\Big)\nonumber\\
   &-&\frac{128 \nu\eta^5 }{567 T^2}+\frac{2^\frac{2}{3}\eta^6}{3^\frac{1}{3}T^\frac{7}{3}}\Big(-\frac{3193 \nu
   ^3}{378}+\frac{3839 \nu
   ^2}{378}\nonumber\\
   &+&\frac{451 \pi ^2 \nu
   }{1512}+\frac{117749 \nu
   }{7938}-\frac{2624}{189}\Big)\nonumber\\
   &+&\frac{2^\frac{1}{3}\eta^7}{3^\frac{2}{3}T^\frac{8}{3}}\Big(\frac{4496 \nu
   ^2}{1701}-\frac{428 \nu }{567}\Big)+O\left(\eta^8\right)\,,
\eea
vanishing both at $T=\pm \infty$.

\section{Nonlocal effects at the 4PN and  4.5PN approximation level} 

The fact that the system loses linear momentum starting from the 3.5PN level of accuracy implies that the cm frame is no more an inertial frame, and one has to face with the presence of inertial forces, i.e., with dragging effects (besides the radiation-reaction effects). For example, if one uses a Hamiltonian description for the conservative part of the problem and adds external forces to represent both radiation-reaction and dragging effects due to inertial forces, ${\mathcal F}_i={\mathcal F}_i^{\rm rr}+{\mathcal F}_i^{\rm drag}$, the dynamics of the system is modified as
\beq
\label{eq_of_moto}
\dot x^i =\frac{\partial H(x,p)}{\partial p_i}\,,\qquad \dot p_i=-\frac{\partial H(x,p)}{\partial x^i}+{\mathcal F}_i\,.
\eeq
Ref. \cite{Blanchet:2026suq} has provided us with the harmonic coordinate expression of ${\mathcal F}_i$  (see Eq. (6.18) and (6.19) there) which start affecting the two-bodies dynamics at 4.5PN, besides a nonlocal contribution to the radiation-reacted dynamics which starts already at 4PN,
\beq
\label{nonloc4PN}
{\mathcal F}_{\rm rr, nl, 4PN}^i =-\frac{4\eta^8 GM^2 \nu }{5} x^j \int_0^\infty d\tau \ln \left(\frac{c\tau}{2P}\right){\mathcal I}^{\rm sym}_{ij}(t,\tau)\,,
\eeq
where
\beq
{\mathcal I}^{\rm sym}_{ij}(t,\tau)=I_{ij}^{(7)}(t-\tau)+I_{ij}^{(7)}(t+\tau)\,. 
\eeq

Specifically, we will evaluate below along the orbit the nonlocal 4PN term, Eq. \eqref{nonloc4PN} (see also Eq. 6.8 in Ref. \cite{Blanchet:2026suq}) as well as
the \lq\lq rad-term" of Eq. (6.10) in Ref. \cite{Blanchet:2026suq} as an example. 
We find
\beq
{\mathcal I}^{\rm sym}_{ij}(t,\tau)=[f_2^{(7)}(t-\tau)+f_2^{(7)}(t+\tau)]q_{ij}\,, 
\eeq
so that 
\bea
\label{nonloc4PNexpl}
{\mathcal F}_{\rm rr, nl, 4PN}^1 &=&-\frac{4928 M^4 \pi \nu^2}{81t^{2/9}}\,, \nonumber\\
{\mathcal F}_{\rm rr, nl, 4PN}^2 &=& 0\,,
\eea
carefully extracting the appropriate n-th roots of unity.
Furthermore, from Eqs. 6.11a and b of Ref. \cite{Blanchet:2026suq}) we have
\bea
\mathcal{F}_{P}^i&=&\eta^7\frac{M^4\nu^2\sqrt{1-4\nu}}{r^4}\Bigg\{\dot r n^i\Big(\frac{32M}{35r}-\frac{24}{7}\dot r^2+\frac{88}{21}v^2\Big)\nonumber\\
&+&v^i\Big(-\frac{64M}{105r}+\frac{304}{105}\dot r^2-\frac{80}{21}v^2\Big)\Bigg\}+O\left(\eta^9\right)\,,\nonumber\\
\mathcal{F}_{G}^i&=&\eta^7\frac{M^3\nu^2\sqrt{1-4\nu}}{r^2}\Bigg\{n^i\Big(\frac{272M}{105r}\dot r^2-\frac{64M}{15r}v^2\nonumber\\
&+&\frac{48}{35}\dot r^2v^2-\frac{32}{35}v^4\Big)+\dot r v^i\Big(\frac{16M}{21r}-\frac{24}{7}\dot r^2\nonumber\\
&+&\frac{24}{7}v^2\Big)\Bigg\}+O\left(\eta^9\right)\,.
\eea
The integrated flux of linear momentum is
\bea
\Pi^1(t)&=&\int_{-\infty}^t dt' \mathcal{F}_{ P}^1=\frac{32\cdot 2^\frac{2}{3} \eta ^9 \sqrt{1-4
   \nu } (\nu -2) \nu ^2 M}{2835
   3^\frac{1}{3} T^\frac{10}{3}}\nonumber\\
   &+&O\left(\eta^{10}\right)\,,
\eea
and $\Pi^2(t)=\Pi^3(t)=0$.
Similarly, let us define the integrated flux of cm position (see Eqs. 6.1b) of Ref. \cite{Blanchet:2026suq})
\bea
\Gamma^i(t) &=& \int_{-\infty}^t dt' \int_{-\infty}^{t'} dt'' {\mathcal F}^i_{ P}(t'')+\int_{-\infty}^t dt' {\mathcal F}^i_{G}(t')\,,\qquad
\eea
where, from Eqs.  6.11 a and b  of \cite{Blanchet:2026suq}, we have

A direct computation shows
\bea
\Gamma^1(t) &=& \frac{992\cdot2^\frac{2}{3} \eta^9
   \sqrt{1-4 \nu } (2-\nu ) \nu ^2
   M^2}{6615 \sqrt[3]{3} T^{7/3}}+O\left(\eta^{10}\right)\,,\nonumber\\
   \Gamma^2(t) &=&\Gamma^3(t) =0\,.
\eea
The radiation reaction contribution to the relative acceleration in the CM frame at 4.5PN \cite{Blanchet:2026suq} is given by
\beq\label{arr}
a_{\rm RR\, 4.5PN\, rad}^i=\frac{\eta^2\Delta}{r^3}\left(2x^iv^j+x^j  v^i\right)\Big[\Pi^j+\mathcal{F}^j_{\rm G}\Big]\,.
\eeq
Since in the present case
\beq
x^i=r (1,0,0),\quad v^i=(\dot r,0,0)=\frac{\dot r}{r} x^i\,,
\eeq
Eq. \eqref{arr} simplifies as
\bea
a_{\rm RR\, 4.5PN\, rad}^i&=&\frac{\eta^2\Delta}{r^2}\left(\mathfrak{a}^i_\Pi+\mathfrak{a}^i_{\rm G}\right)\,,\nonumber\\
\eea
with
\bea
\mathfrak{a}^i_\Pi&=& 
=3\frac{\dot r}{r^2} (x\cdot\Pi) x^i\,, \nonumber\\
\mathfrak{a}^i_{\rm G}&{=}&
3\frac{\dot r}{r^2} (x\cdot\mathcal{F}_{\rm G}) x^i\,.
\eea
A direct evaluation gives
\bea
a_{\rm RR\, 4.5PN\, rad}^1&{=}&\eta^{11}\frac{3968M^4}{8505t^5}(1-4\nu)(\nu-2)\nu^2\nonumber\\
&+&O\left(\eta^{12}\right)\,,
\eea
i.e., it is still zero at 4.5PN and starts contributing only at 5.5PN.

\section{Concluding remarks}

We have applied the PN-matched Multipolar Post Minkowskian formalism to the case of a two-body system in radial fall (also termed as a head-on collision, or plunge-in situation). We have computed, within the post-Newtonian (PN) approximation, the corresponding waveform, up to the 3.5 PN accuracy level, updating previous results. At this level the presence of a radiation-reaction force manifests with two contributions: at 2.5PN order (leading-order contribution) and  at 3.5PN order (next-to-leading-order contribution),  modifying the fall with a corresponding bremsstrahlung radiation. 

We have evaluated  all the emissions: energy, angular momentum (vanishing identically) and linear momentum. In addition, we have also evaluated the cm motion and the (nonlocal)  contributions appearing at 4PN (tail) and at 4.5PN (radiation-reaction) due to all these losses, paving the way for future more accurate computations (note that for the waveform the complete 4PN accuracy is not available from existing literature, see e.g. Ref. \cite{Blanchet:2013haa} except for the quasi-circular orbit case). An associated ancillary file contains the explicit expressions for all the main accomplishments of the present work \cite{anc}.

Our analytic results are also validated by comparison with existing numerical results \cite{Detweiler:1979xr}: we have found 
a nice agreement in a region of the frequency domain which is safe from both numerical problems (i.e., avoiding small frequencies, where one needs much high numerical precision) and analytical problems (i.e., avoiding high frequencies, where the PN approximation breaks down).

As already stated in the Introduction (and discussed in previous papers), the PN expansion for a radially infalling particle does not provide access to the full solution of the problem. Indeed, by definition, the PN expansion can reconstruct the relevant observables only in the weak field regime, i.e., far from the BH horizon (at the horizon the radial fall terminates by a capture process, and  the particle inevitably probes the strong-field region). For this reason, the present work is also preliminary to future, forthcoming analytical studies of the same problem directly in the strong-field regime.

\section*{Acknowledgments}

D.B. and G.D.R. thank S. Albanesi, A. Cipriani, A. Geralico, A. Nagar for fruitful discussions.
D.B. and G.D.R. acknowledge  membership to the Italian Gruppo Nazionale per
la Fisica Matematica (GNFM) of the Istituto Nazionale
di Alta Matematica (INDAM).

\bibliographystyle{apsrev}
\bibliography{references}

\end{document}